\documentclass[a4paper,12pt,reqno,superscriptaddress,nofootinbib]{revtex4}
\usepackage[centertags]{amsmath}
\usepackage{amsfonts}
\usepackage{amssymb}
\usepackage{amsthm}
\usepackage{newlfont}
\usepackage{stmaryrd}
\usepackage{mathrsfs}
\usepackage{mathtools}
\usepackage{euscript}
\usepackage{graphicx}
\usepackage{enumerate}
\usepackage{todonotes}

\usepackage{color}

\usepackage{tikz}
\usepackage{pgf}
\usetikzlibrary{positioning,fit,calc}
\usetikzlibrary{arrows,automata}
\usepackage{wrapfig}
\usepackage{subfigure}
\usepackage{amscd}
\usepackage{hyperref}

\usepackage{changes}

\theoremstyle{plain}

\theoremstyle{definition}

\theoremstyle{remark}

\let\de=\delta 
\let\ve=\varepsilon   
  \let\om=\omega 
   
\let\ze=\zeta 

   \let\Om=\Omega

\newcommand{\be}{\begin{equation}}
\newcommand{\en}{\end{equation}}

\def\om{\omega}

\let\lam=\lambda
   
\let\ze=\zeta 

   \let\Om=\Omega
 
\let\om=\omega

\newcommand{\caF}{{\mathcal F}}

\newcommand{\caS}{{\mathcal S}}

\newcommand{\opunit}{\text{1}\kern-0.22em\text{l}}

\newcommand{\id}{\textrm{d}}

\DeclareMathAlphabet{\mathpzc}{OT1}{pzc}{m}{it}

\newcommand{\rel}{\,|\,}

\def\dbar{{\mathchar'26\mkern-12mu d}}

\let\oldsqrt\sqrt
\def\sqrt{\mathpalette\DHLhksqrt}
\def\DHLhksqrt#1#2{%
	\setbox0=\hbox{$#1\oldsqrt{#2\,}$}\dimen0=\ht0
	\advance\dimen0-0.2\ht0
	\setbox2=\hbox{\vrule height\ht0 depth -\dimen0}%
	{\box0\lower0.4pt\box2}}

\let\de=\delta 
\let\ve=\varepsilon   
  \let\om=\omega 
   
\let\ze=\zeta 
\let\be=\beta

   \let\Om=\Omega

\DeclareMathAlphabet{\mathpzc}{OT1}{pzc}{m}{it}

\def\bea{\begin{eqnarray}}
\def\eea{\end{eqnarray}}
\def\ba{\begin{array}}
	\def\ea{\end{array}}

\begin{document}

\title{Nonequilibrium corrections to gradient flow}

\author{Christian Maes}
\affiliation{Instituut voor Theoretische Fysica, KU Leuven, 3001 Leuven, Belgium}
\author{Karel Neto\v{c}n\'{y}}
\email{netocny@fzu.cz}
\affiliation{Institute of Physics, Czech Academy of Sciences, Prague, Czech Republic}

\begin{abstract}
The force on a probe induced by a nonequilibrium medium is in general nongradient.  We detail the mechanism of that feature via nonequilibrium response theory.  The emergence of nongradient forces is due to a systematic ``twist'' of the excess frenesy with respect to the entropy flux, in response to changes in the coupling or in the position of the probe in the nonequilibrium medium.
\end{abstract}
\maketitle

\noindent {\bf Stationary nonequilibria are found in a wide range of natural phenomena, from space plasmas to cell life.  When a slow probe is immersed in and coupled to a nonequilibrium medium, it will experience a systematic force.  That induced force need not be derivable from a potential, in contrast with the situation for equilibrium media where the forces are gradients of thermodynamic potentials. That implies that the probe dynamics may be much richer than under thermodynamic equilibrium.  This paper investigates what exactly causes that nongradient effect, and how it can be quantified.  For that, it is useful and natural to use response theory for nonequilibrium systems.}

\baselineskip=20pt
{\large
\section{Introduction}
Nonequilibrium statistical mechanics has evolved much since its beginning in irreversible thermodynamics \cite{deg,ons}.  Not only steady nonequilibrium baths but also small systems for which no notion of local equilibrium makes sense have become important research topics. The motivation arises from a wealth of new data ranging from atmosphere dynamics over micro-rheology and polymer science to biological physics. One question of considerable interest is to characterize the mean force on a particle induced by some nonequilibrium environment.  Let us explain the various ingredients in that question.  The particle will be characterized by a position $x$ in some spatial domain, which is coupled via an interaction potential $U(x,\eta)$ to much faster degrees of freedom $\eta$ in the environment.  The $\eta$ evolve in contact with a thermal bath but are also subject to some driving which we leave unspecified for now. Generally, the $\eta$-dynamics is assumed given by a Markov stochastic evolution in which the nonequilibrium is visible from time-reversal breaking in the steady condition.  As the $\eta$ may sometimes also be positions of (other) particles, in order to avoid confusion, we will call the particle (with position $x$) under consideration the {\it probe}.  One should indeed imagine a somewhat heavier or bigger and at any rate, much slower particle for that probe. We are then entering the regime of the so called quasistatic limit in which the dynamics of the $\eta$ will only be considered at fixed position $x$ of the probe. What matters for the mean force on the probe is their stationary law $\rho_x$ at fixed $x$, which is a stationary density to be solved from the appropriate Smoluchowski, Fokker-Planck or Master equation.  We define then the mean force to be
\begin{equation}\label{gred}
f(x) :=  -\langle \nabla_x U(x,\eta)\rangle_x= -\int \,\nabla_x U(x,\eta)\,\rho_x(\id\eta)
\end{equation}
with $\nabla_x$ the gradient of the position $x$.
The integral or the expectation $\langle\cdot\rangle_x$ is over the $\eta$, while fixing $x$.\\

 Later will be given more details on what type of $\eta-$dynamics is specifying the $\rho_x$.  Note however that the above scheme leads to a so called gradient force in the case of a reversible dynamics for positions $\eta$ in equilibrium.  The reason is that the probe is in contact then with a canonical equilibrium at inverse temperature $\beta = (k_BT)^{-1}$, leading for \eqref{gred} to a mean force,
\begin{eqnarray}\label{gr}
f_\text{eq}(x) &=& -\frac 1{Z(x,T)}\int\id \eta \,\nabla_x U(x,\eta)\,e^{-\beta U(x,\eta)}\nonumber\\
&=& -\nabla_x {\cal F}_\text{eq}(x),\qquad {\cal F}_\text{eq}(x) :=-k_BT \log Z(x,T)
\end{eqnarray}
where $Z(x,T)$ is the canonical partition function.
As that  mean force \eqref{gr} is gradient, the work
\begin{equation}\label{wor}
\int_{\gamma:x_\text{i}\rightarrow x_\text{f}}\, f_\text{eq}(y)\cdot \id y = {\cal F}_\text{eq}(x_\text{i}) - {\cal F}_\text{eq}(x_\text{f})
\end{equation}
just depends on the final and initial positions $x_\text{f}, x_\text{i}$ in the probe trajectory $\gamma$.  For isothermal changes as above, that reproduces the reversible expression for the work done to be  the difference in free energy ${\cal F}_\text{eq}$.\\

There are generalizations of the above question for characterizing \eqref{gred} (and of the results to come) when the probe is not quite a particle or is a more abstract entity, e.g. representing a container wall holding a fluid or being a slow collective variable defined from the $\eta$ degrees of freedom.  For example, the gradient structure of the force illustrated in \eqref{gr} is relevant for a great number of thermodynamic relations in equilibrium and for the very existence of certain state functions \cite{cal}.  It is also at the basis of characterizing the dissipative relaxation to equilibrium as a gradient flow \cite{vil,ot}.  That ``mechanical'' picture breaks down for quasistatic transformations in contact with nonequilibria \cite{ru,knst}.  The present paper gives general characterizations under which the mean force \eqref{gred} becomes nongradient, and what physics (and quantities) determine that.\\

The main ingredient of the analysis is nonequilibrium response theory, here applied to Markov processes and explained in Section \ref{resp}; see also \cite{fdr,njp}.   The general point is to start from a reference process $P_\text{ref}$ and consider its $\log$ Radon-Nikodym derivative (called action $\cal A$) with respect to the process $P$ under investigation. There is a finite time-window over which we assume absolute continuity.  We then obtain from the Girsanov theorem \cite{girs},
\begin{equation}\label{rnd}
\frac{\id P}{\id P_\text{ref}} = e^{-\cal A}, \qquad \cal A = D - S/2
\end{equation}
where, by definition,  $S$ is the antisymmetric part and $D$ is the time-symmetric part in the action $\cal A$.  Quite informally, it means that the weight of an $\eta$-trajectory $\omega = (\eta_s, 0\leq s\leq t)$ in time $[0,t]$ is given by
\begin{equation}\label{beg}
\text{Prob}[\omega] \propto e^{S(\omega)/2 -D(\omega)}
\end{equation}
with $S$ being antisymmetric and $D$ being symmetric under time-reversal.   We have ignored the parameter-dependencies in the notation, in particular how $S$ and $D$ depend on the probe position $x$ and on the coupling between probe and environment.  That will be made explicit in Section \ref{resp} as formula \eqref{rnd} is the start of a systematic response theory.  That theory is helped substantially by the physical interpretation of $S$ and $D$.   In the context of the present paper $S$ can be interpreted as the (physical) entropy production in the process $P$ in excess with respect to the reference process $P_\text{ref}$.  That $S$ is indeed the (physical) entropy production is not automatic but holds under a condition which is known as {\it local detailed balance} and also assumes a sufficiently weak coupling of the $\eta$ with respect to the equilibrium reservoir in which heat is dissipated, \cite{time}.  The variable $D$ is called the {\it frenesy} (by definition, the time-symmetric part in the action) and collects kinetic information related to escape rates and dynamical activity \cite{fren}.  See Appendix \ref{dac} for the expression of $D$ in the case of Markov jump processes.\\

   We start with
 the presentation of a toy-model in the next section. It illustrates what we are after.  The necessary response theory for a more general understanding is reviewed in Section \ref{resp}. We study the problem in a weak coupling expansion in Section \ref{weak}.  It is the nontrivial dependence of the frenesy on the coupling that makes the statistical force nongradient.  Similarly, the change in frenesy of the medium by moving the probe position is proportional to the rotational component in the statistical force.  In fact, in Section \ref{stif} we show that the curl of the statistical force is given in terms of an external product between entropy flux and excess frenesy.
     When the medium is in equilibrium, these rotational components disappear of course.  Finally, in Section \ref{notso} we investigate the appearance of nongradient forces up to second order around equilibrium, comparing also the response theory of Section \ref{resp} with an approach introduced by Komatsu and Nakagama~\cite{naokom}.

\section{An example}\label{toy}
Let us start with a toy model to illustrate in a simple setting the main phenomenon.  The paper is not about a specific model  but the following is useful because it allows an explicit calculation.\\
We have a probe with positions on the ring, $x \in S^1$, and in interaction with a ``spin'' degree of freedom $\eta = 0, 1$.   The interaction between $x$ and $\eta$ goes via potential $U(x,\eta)$.  There is also a thermal bath represented by its inverse temperature $\beta\neq 0$.  The dynamics of $\eta$ is spin-flip, following a Markov jump process with Master equation
\begin{eqnarray}\label{me}
  \frac{\partial}{\partial t}\rho_x(\eta,t) &=&
  [k_R(1-\eta,\eta) + k_L(1-\eta,\eta)]\,\rho_x(1-\eta,t)\nonumber\\
 && -[k_R(\eta,1-\eta) + k_L(\eta,1-\eta)]\,\rho_x(\eta,t)
  \end{eqnarray}
at fixed probe position $x$, for the time-dependent probability $\rho_x(\eta,t)$ of $\eta$.  For the transitions we see two channels, left (L) and right (R), by which
$0 \stackrel{L,R}{\longleftrightarrow} 1$.  The transition rates have the general form,
\begin{align}
k_{L,R}(0,1) &= a_{L,R}(x)\,e^{-\frac{\be}{2}[u(x) \pm \ve]}
\\
k_{L,R}(1,0) &= a_{L,R}(x)\,e^{\frac{\be}{2}[u(x) \pm \ve]}\,,
\qquad u(x) = U(x,1) - U(x,0)\nonumber
\end{align}
 The parameter
$\ve$ reads the work of driving forces along the cycle
$0 \stackrel{R}{\rightarrow} 1 \stackrel{L}{\rightarrow} 0$.  Note that not only the energy $U(x,\eta)$ but also the kinetic factors
$a_{L,R}(x)$ depend on the position of the probe which makes the coupling between $x$ and $\eta$.\\
  From \eqref{me} it is trivial to find the stationary occupations.  They satisfy
\begin{equation}\label{bzet}
\frac{\rho_x(1)}{\rho_x(0)} =
\frac{k_L(0,1) + k_R(0,1)}{k_L(1,0) + k_R(1,0)} =
\ze(x)\,e^{-\be u(x)}
\end{equation}
where
\begin{equation}\label{azet}
\ze(x) = \frac{a_R(x) + a_L(x)\,e^{-\be\ve}}{a_L(x) + a_R(x)\,e^{-\be\ve}}.
\end{equation}
The statistical force $f(x)$ on the probe as defined in \eqref{gred} is the mean force in the quasistatic limit where the probe is fixed at position $x$: here,
\begin{equation}\label{1f}
f(x)= -U'(x,1) \,\rho_x(1) - U'(x,0) \rho_x(0).
\end{equation}
Hence, inserting \eqref{bzet}--\eqref{azet}, the statistical force \eqref{1f} on the probe equals
\begin{equation}\label{2level-force}
f(x)
= -U'(x,0) - \frac{u'(x)}{1 + \ze^{-1}(x)\,e^{\be u(x)}},\,\qquad x\in S^1
\end{equation}
The question of the paper can be illustrated here by asking when and when not that force $f(x), x\in S^1$, has a rotational part.  I.e., when is $\oint f(x)\id x \neq 0$?\\

It is easy to check that the statistical force is derived from the free energy
\[
f(x) = \frac 1{\beta} \frac{\id}{\id x} \log \big( e^{-\beta U(x,0)} +e^{-\beta U(x,0)}\big)\quad \text{ whenever } \ze(x) \equiv 1
\]
(which is the usual equilibrium formula for the statistical force) when
$\ve = 0$ (detailed balance) or for $a_L = a_R$ (channel symmetry). Those are however not the only cases here for which the statistical force is a gradient.  Indeed, when the kinetic factors in \eqref{azet} are ``energetic'' in the sense that they depend on the position $x$ entirely via the energy gap, i.e., $\ze(x) = z(u(x))$ for some function $z(\cdot)$, then
the statistical force is also gradient,
$f(x) = -\tilde\caF'(x)$ with respect to the ``free energy'' (up to any constant),
\[
\tilde\caF(x) = U(x,0) + \int^{u(x)} \frac{\id v}{1 + z^{-1}(v)\,e^{\be v}}
\]
Therefore to obtain nongradient forces it is necessary that, loosely speaking, the positions with the same energy gap are ``resolved'' via distinct kinetic factors.

In order to find a sufficient condition for a nongradient contribution  we can be more explicit via an expansion in a small probe-medium coupling constant $\lambda$,
\[
u(x) = u_0 + \lambda u_1(x) + \ldots \text{  and }
\ze(x) = \ze_0 + \lambda \ze_1(x) + \ldots
\]
To second order in the coupling strength $\lambda$, the formula~\eqref{2level-force} yields
\begin{equation}
\oint f\,\id x = \frac{\lambda^2}{(e^{\be u_0 /2} + \ze_0\, e^{-\be u_0 /2})^2}
\oint u_1(x)\,\ze'_1(x)\,\id x + O(\lambda^3).
\end{equation}
For example, if the kinetic factor of the L-channel is harmonically modulated around a symmetric uncoupled reference,
\[ 
a_L(x) = a + \lambda\,b\, \cos(2\pi x),\qquad a_R(x) \equiv a>0
\] 
then the nonequilibrium factor reads
\[ 
\ze(x) = 1 - \lambda\,\frac{b}{a}\tanh \bigl( \frac{\be\ve}{2} \bigr)
\cos(2\pi x) + O(\lambda^2)
\] 
and assuming a (shifted) harmonic modulation also in the level spacing,
\[ 
u(x) = u_0 + \lambda \hat u_1 \cos (2\pi x - \phi),\qquad \hat u_1\neq 0
\] 
the nongradient part in the force equals
\begin{equation}
\oint f\,\id x = \lambda^2\,
\frac{ \hat u_1 \,b}{8a\, \cosh^2 \bigl( \frac{\be u_0}{2} \bigr)} \tanh \bigl( \frac{\be\ve}{2} \bigr) \sin\phi
+ O(\lambda^3)
\end{equation}
We observe that for a nonzero gradient force we need not only that the (fast) two-level system is driven ($\ve\neq 0$) or the channel is asymmetric ($b\neq 0$), but also that there is a phase shift (or ``twist'') ($\phi\neq 0$) between the spatial modulations in the level spacing and in the kinetic factor, respectively.  We will see that effect again and discussed more generally in the following section.

\section{Response out-of-equilibrium}\label{resp}
We compare two $\eta-$processes, $P_1$ and $P_2$, both corresponding to an $\eta$-dynamics but for different coupling $\lambda$ between probe and environment $\eta$, or for different (fixed) probe position $x$. Compared to the discussion around \eqref{rnd} we take the same reference process $P_\text{ref}$ for both, while $P=P_1$ or $P=P_2$.  The initial state at time zero is supposed to be the same for both.  The processes can be compared via expectations of observables $f(\omega)$ on trajectories $\omega = [\eta_s]_0^t$,
\[
\langle f(\omega)\rangle_1
= \int f(\omega)\,\id P_1(\omega)\;\;\text{ versus }\;\; \langle f(\omega)\rangle_2
= \int f(\omega)\,\id P_2(\omega)
\]
and $\id P_2 =  \id P_1\,e^{-{\cal A}}$ where as in \eqref{rnd}, the action
\[
{\cal A}(\omega) = D_2(\omega) -D_1(\omega) - \frac 1{2} [S_2(\omega) - S_1(\omega)]
\]
is
expressed in changes of frenesy, and changes in entropy flux.\\

Let us apply that now in the spirit of \cite{naokom} for the probability to find state $\eta$ at time $t$. We choose $f(\omega)= \delta(\eta_t-\eta)$,
\begin{equation}\label{eg}
p^t_{2}(\eta) = \langle \delta(\eta_t-\eta)\rangle_2 = \int e^{-{\cal A}(\omega)}\, \delta(\eta_t-\eta)\,\id P_1(\omega)
\end{equation}
and dividing that by $\langle \delta(\eta_t-\eta)\rangle_1$ we get for all times $t$, including those far beyond the relaxation time of the medium, that
\begin{eqnarray}\label{ppt}
\frac{p^t_{2}(\eta)}{p^t_{1}(\eta)} &=& \langle e^{-{\cal A}(\omega)}\, |\,\eta_t=\eta\rangle_1\nonumber\\
&=& \langle e^{D_1(\omega)-D_2(\omega)  + \frac 1{2} [S_2(\omega) - S_1(\omega)]}\, |\,\eta_t=\eta\rangle_1
\end{eqnarray}
For a small change from the first to the second process we obviously need the derivatives of $D$ and $S$ with respect to coupling $\lambda$ and probe position $x$, to write
\[
e^{-{\cal A}(\omega)} =  1- (\lambda_2-\lambda_1)\left[\frac{\partial D}{\partial \lambda}  - \frac 1{2}\frac{\partial S}{\partial \lambda}\right] - (x_2- x_1)\cdot\nabla_x \bigl(
D- \frac{S}{2} \bigr) + \ldots
\]
where the $[\cdot]$ must be evaluated at $\lambda_1,x_1$.  That can be plugged into \eqref{ppt} for obtaining the small changes in the statistics at time $t$.  

At the same time, e.g. from \eqref{eg} by summing over $\eta$ we always have the normalization
\[
\int e^{-{\cal A}(\omega)}\,\,\id P_1(\omega) = 1
\]
and hence
\begin{equation}\label{norms}
\Bigl\langle \frac{\partial D}{\partial \lambda} \Bigr\rangle_1  =  \frac 1{2}\Bigl\langle\frac{\partial S}{\partial \lambda}\Bigr\rangle_1, \quad \Bigl\langle\nabla_x \bigl( D- \frac{S}{2} \bigr) \Bigr\rangle_1 = 0
\end{equation}
for no matter what initial condition.

\section{Weak coupling expansion}\label{weak}

Assume that the coupling potential between probe and medium has the form
$U_\lam (x,\eta) = U_0(\eta) + \lam U_I(x,\eta)$. Expanding the force \eqref{gred} in the coupling parameter $\lam $, the leading term
is of gradient form,
$f(x) = -\lam\nabla_x \langle U_I \rangle^{\lam= 0} + O(\lam ^2)$.   The result of this section is an expression, Eq. \eqref{20}, for the second order $O(\lam ^2)$ in the weak coupling expansion, where we will see the appearance of a rotational part in the force.  That is the relevant term in a weak coupling limit in the spirit of \cite{vanh}. Note that as in such Van Hove limits we will not need a detailed or specific model dynamics for the environment.\\

The method is the response theory of Section \ref{resp}. It is already clear from there that, in contrast with equilibrium, we need here also the excess in frenesy $D$, defined for each $\eta$-trajectory, and that will make the interesting difference.  

Before we proceed we make one extra physical assumption.  We have seen in the previous paragraph that we need the change in entropy flux when modifying the coupling.  Here we assume that
\begin{equation}\label{as}
S'(\om) = \be\,[U_I(x,\eta_0) - U_I(x,\eta_t)]
\end{equation}
 with the shorthand
$' \equiv \id / \id \lam|_{\lam=0}$.  Physically that means to suppose that the probe does not directly interfere with the nonequilibrium driving on the $\eta$-medium in the sense that the (path-wise) work $W(\om)$ of driving forces
 is independent of $\lam $ or $x$. Here,  $\om = [\eta_s]_0^t$ denotes the trajectory in the nonequilibrium medium and the entropy flux per $k_B$ is
\begin{equation}\label{sho}
S(\om) = \be\,[U_\lam (x,\eta_0) - U_\lam (x,\eta_t) + W(\om)].
\end{equation}
That is obviously relevant for the response to changes in the coupling.\\

We now use formula \eqref{ppt} with process $P_2$ corresponding to coupling $\lambda$ and process $P_1$ with coupling $\lambda=0$. We start both processes from the stationary distribution $\rho^0$ of the medium for the uncoupled case.  The stationary distribution at coupling $\lambda$ can then be evaluated as in \eqref{ppt}, for $t \to \infty$,
\begin{equation}\label{spt}
\begin{split}
\frac{\id\rho^\lam _x}{\id\rho^0}(\eta) &=
\Bigl\langle e^{(S^\lam- S^0)/2 - (D^\lam- D^0)}
\,\Bigl|\,\Bigr. \eta_t = \eta
\Bigr\rangle^0
\\
&= 1 + \lam \, \Bigl\langle \frac{S'}{2} - D'
\,\Bigl|\,\Bigr. \eta_t = \eta \Bigr\rangle^0 -
\lam \, \Bigl\langle \frac{S'}{2} - D'
\,\Bigl|\,\Bigr. \eta_0 = \eta \Bigr\rangle^0
+ O(\lam ^2)
\\
&= 1 - \lam \be\,[ U_I(x,\eta) - \langle U_I \rangle^0 ] + \lam H'_x(\eta) + O(\lam^2).
\end{split}
\end{equation}
We have used relations \eqref{norms}--\eqref{as} for adding $\Bigl\langle \frac{S'}{2} - D'
\,\Bigl|\,\Bigr. \eta_0 = \eta \Bigr\rangle^0=0$, and that $\lim_t \langle U_I(x,\eta_0)\,|\,\eta_t =\eta\rangle^0 = \langle U_I \rangle^0$. We denote
\[
H'_x(\eta)
= \lim_{t\uparrow \infty} \bigl[ \langle D'_x \rel \eta_0 = \eta \rangle^0 -
\langle D'_x \rel \eta_t = \eta \rangle^0 \bigr].
\]
Per consequence the force on the probe is
\begin{eqnarray}\label{f1}
f_\lam (x) &=& -\lam \, \Bigr\langle \frac{\id\rho^\lam _x}{\id\rho^0}\,
\nabla_x U_I \Bigr\rangle^0\nonumber\\
&=& -\lam\nabla_x \langle U_I \rangle^0 + \frac{\lam ^2 \be}{2}
\nabla_x \bigl[ \langle U_I^2 \rangle^0 -
\bigl( \langle U_I \rangle^0 \bigr)^2 \bigr] -
\lam ^2 \langle H'\,\nabla_x U_I \rangle^0 + O(\lam ^3)\nonumber\\
&=& -\nabla_x \Psi_\lam (x) + f_\lam^\text{neq}(x)
\end{eqnarray}
with the potential defined in the decoupled medium,
\begin{equation}
\begin{split}
\Psi_\lam (x) &= -\frac{1}{\be}\log\,
\bigl\langle e^{-\be\lam U_I}\bigr\rangle^0
\end{split}
\end{equation}
which obviously reduces to the equilibrium free energy (difference) under detailed balance, and the second contribution is generally nonconservative,
\begin{equation}\label{nonconserv}
\begin{split}
f_\lam ^\text{neq}(x) &= -\lam ^2 \langle H'\,\nabla_x U_I \rangle^0 + O(\lam ^3).
\\
&=
-\frac{1}{\be} \bigl\langle (D_\lam - D_0)\,\nabla_x (S_\lam - S_0)
\bigr\rangle^0 + O(\lam ^3).
\end{split}
\end{equation}
As suggested by the last (formal) expression the latter originates from correlations between the (gradient of) entropy flux and the frenesy. (Note that the $S_0$ is redundant there.)

The decomposition of the induced force into conservative and nonconservative components is of course not unique (unless imposing some extra condition like, e.g., that the latter component is to be divergence-free as in the Helmholtz decomposition). An alternative representation would be, for example,
\begin{equation}\label{f2}
f_\lam (x) = -\nabla_x \tilde\Psi_\lam (x) + \tilde f_\lam ^\text{neq}(x)
\end{equation}
with the modified
potential (now in the coupled medium)
\begin{equation}
\begin{split}
\tilde\Psi_\lam (x) &= \frac{1}{\be}
\log\, \bigl\langle e^{\be\lam U_I} \bigr\rangle^\lam
\end{split}
\end{equation}
and the modified nonconservative force
\begin{equation}
\begin{split}
\tilde f_\lam ^\text{neq}(x) &= \lam ^2\langle U_I \nabla_x H' \rangle^0
+ O(\lam ^3)
\\
&= \frac{1}{\be} \bigl\langle (S_\lam - S_0)\,\nabla_x (D_\lam - D_0)
\bigr\rangle^0 + O(\lam ^3)
\end{split}
\end{equation}
We see that it is the position dependence of the medium's frenesy that delivers the nongradient contribution.

The above is easily checked by using the response formula \eqref{ppt} or \eqref{spt} directly for the averaged interaction potential
\begin{equation}
\langle U_I \rangle^\lam =
\langle U_I \rangle^0 - \lam \be\,\bigl[ \langle U_I^2 \rangle^0 -
\bigl( \langle U_I \rangle^0 \bigr)^2 \bigr] + \lam \langle H'\,U_I \rangle^0
+ O(\lam ^2).
\end{equation}
Finally, observe that by adding the two expressions \eqref{f1} and \eqref{f2} we get the induced force as
\begin{eqnarray}\label{20}
2\beta\,f_\lambda(x) &=& -\nabla_x
\log\, \frac{\bigl\langle e^{\be\lam U_I} \bigr\rangle^\lam}{\bigl\langle e^{-\be\lam U_I}\bigr\rangle^0}\\
&+& \bigl\langle (S_\lam - S_0)\,\nabla_x (D_\lam - D_0) - (D_\lam - D_0)\,\nabla_x (S_\lam - S_0)
\bigr\rangle^0 + O(\lam ^3),\nonumber
\end{eqnarray}
where the nongradient part (in the second line) contains the ``bracket'' $\partial_\lambda S\;\nabla_x D - \partial_\lambda D\;\nabla_x S$.  We see that the mean force has acquired a rotational part by the presence of the frenesy $D_\lambda$ when that kinetic contribution is ``twisted'' with respect to the thermodyanmic information contained in the entropy flux $S_\lambda$.

\section{Stiffness}\label{stif}
We can also apply the linear response formalism of Section \ref{resp} to establish local properties of the mean force \eqref{gred} in the ``thermodynamic space'' of probe coordinates
$x = (x_1,\ldots,x_d)$. More specifically, we can study the (differential) stiffness of the probe in terms of the linear response matrix
\begin{equation}\label{43}
M_{jk}(x) = -\nabla_j f_k(x)
= \nabla_j \langle \nabla_k U \rangle_x\,,\qquad
\nabla_j \equiv \frac{\partial}{\partial x_j}.
\end{equation}
We remind that $\langle\cdot\rangle_x$ denotes expectation over the nonequilibrium medium with probe fixed at $x$.
Note also that under equilibrium conditions and as a continuation from \eqref{gr} that matrix is
\begin{equation}\label{eqsti}
M^\text{eq}_{jk}(x) =
-k_BT\,\partial_{jk} \log Z(x) =
\langle \partial_{jk} U \rangle_x^\text{eq} -
\be\,\text{Cov}(\nabla_j U; \nabla_k U)_x^\text{eq},
\end{equation}
where the first term represents an inherent (mechanical) stiffness of the probe, which is diminished by fluctuations in the medium (the second term).
It always satisfies the Maxwell symmetry relations
$M^\text{eq}_{jk}(x) = M^\text{eq}_{kj}(x)$, obviously  a direct consequence of the existence of a potential for the thermodynamic force.

Out of equilibrium the matrix $M(x)$ generally decomposes into a symmetric part,
$M^{(s)}(x)$, corresponding to locally conservative forces, and an antisymmetric part,
$M^{(a)}(x)$, representing rotational forces.\\
The differential stiffness \eqref{43} is easily computed via noticing
\begin{equation}\label{ls}
M_{jk}(x) = \langle \partial_{jk} U \rangle_x +
\langle \nabla_j\log\rho \, \nabla_kU \rangle_x
\end{equation}
where $\rho=\rho_x$ is the stationary distribution of the medium.  We apply now the response formula \eqref{ppt} for fixed coupling $\lambda$ and with initial condition given by $\rho_x$.  For $t \to \infty$,
\begin{equation} \label{sip}
\begin{split}
\rho_{x+\id x}(\eta) &= \bigl\langle \de(\eta_t - \eta)
\bigr\rangle_{x+\id x}
\\
&= \rho_x(\eta)\,\Bigl\{ 1 +
\Bigl\langle
\frac{\id S_x}{2} - \id D_x \,\bigl|\, \bigr.
\eta_t = \eta \Bigr\rangle_x
- \Bigl\langle
\frac{\id S_x}{2} - \id D_x \,\bigl|\, \bigr.
\eta_0 = \eta \Bigr\rangle_x
\Bigr\}
\\
&= \rho_x(\eta)\,\bigl\{
1 - \be\,[ \id U(x,\eta) - \langle \id U \rangle_x] + \dbar H(x,\eta) \bigr\}
\end{split}
\end{equation}
where $\dbar H(x,\eta) := \lim_t \langle \id D_x \rel \eta_0 = \eta \rangle_x -
\langle \id D_x \rel \eta_t = \eta \rangle_x$,
and we have again used \eqref{norms} for $\Bigl\langle
\frac{\id S_x}{2} - \id D_x \,\bigl|\, \bigr.
\eta_0 = \eta \Bigr\rangle_x=0$ and
$\id S_x(\om) = \be\,[\id U(x,\eta_0) - \id U(x,\eta_t)]$ due to a variation of the assumption \eqref{as} that the driving contribution to the entropy flux does not (explicitly) depend on $x$.
Equivalently,
\begin{equation}
\id\log\rho_x(\eta) =
- \be\,[ \id U(x,\eta) - \langle \id U \rangle_x] + \dbar H(x,\eta).
\end{equation}
We can insert that in \eqref{ls}.  Writing $\dbar H(x,\eta) = H_j\,\id x_j$, the differential stiffness \eqref{43} equals
\begin{equation}
M_{jk}(x) 
= \langle \partial_{jk} U \rangle_x -
\be\,\text{Cov}(\nabla_j U;\,\nabla_k U)_x +
\text{Cov}(\nabla_k U;\, H_j)_x
\end{equation}
Comparing that with \eqref{eqsti} we see that the last term makes the typical nonequilibrium contribution.  It can equivalently be written as
\begin{equation}\label{stiff-noneq}
\begin{split}
\text{Cov}(\nabla_j U;\, H_k)_x &=
\bigl\langle \bigl[ \nabla_j U(x,\eta_0) - \nabla_j U(x,\eta_\tau) \bigr]\,
\nabla_k D \bigr\rangle_x
\\
& = k_BT\, 
\bigl\langle \nabla_j S\,\nabla_k D \bigr\rangle_x.
\end{split}
\end{equation}
In particular, the antisymmetric part violating the Maxwell relations is, formally,
\begin{equation}
M^{(a)}_{jk}(x) = k_BT 
\,\bigl\langle
\nabla_j S\,\nabla_k D - \nabla_k S\,\nabla_j D \bigr\rangle_x,
\end{equation}
which can also be expressed more elegantly by employing the external form calculus (with $X \wedge Y = -Y \wedge X$ and $\id^2 = 0$) in the form
\begin{equation}
\begin{split}
\id\langle\id U \rangle^x &= M^{(a)}_{jk} \id x_j \wedge \id x_k
\\
&= k_BT \,
\langle \id S \wedge \id D \rangle_x.
\end{split}
\end{equation}
Note that
$\id S \wedge\id D$ is the volume form of a parallelogram demarcated by (co-)vectors
$\id S$ and $\id D$ (pathwise).

\section{Not-so-close-to-equilibrium regime}\label{notso}
Statistical forces can also be investigated in an expansion near equilibrium. Via the condition of local detailed balance it is mostly possible to speak about a driving amplitude $\ve$, as present in the work $W$ in \eqref{sho} done by the driving.  Linear order (in $\ve$) corrections have been given in \cite{prl}.  Here we add the second order correction, for which we employ the Komatsu-Nakagawa formula, \cite{naokom}, and which, to be self-contained, is derived in Appendix \ref{knv} from the response formalism of Section \ref{resp}.

The expression for the stationary distribution obtained in \cite{naokom} (formula \ref{qual} in Appendix \eqref{knv}) reads
\begin{equation}\label{naoko}
\log\frac{\id\rho}{\id\rho^\text{eq}}(\eta) = \be\Om -
\frac{1}{2}\bigl[ \langle S_\text{i} \rangle_\eta -
\langle S_\text{i} \rangle_{\rightarrow \eta} \bigr] + O(\ve^3)
\end{equation}
with $\ve$ the driving amplitude and where the difference in the square bracket reads the difference between the irreversible entropy production $S_i = \beta W$ (linear in $\ve$) along the process started from $\eta$ and the process conditioned on ending at $\eta$. The expectations are denoted by $\langle\cdot\rangle_\eta$, respectively $\langle\cdot\rangle_{\rightarrow \eta}$, and the limit $t\uparrow \infty$ is understood. Furthermore,
\begin{equation}
\be\Om = \caS(\rho \rel \rho^\text{eq}) =
\be\,[\caF(\rho) - \caF(\rho^\text{eq})] + O(\ve^3)
\end{equation}
which is the difference between the ``nonequilibrium free energy''
$\caF^\text{neq} := \caF(\rho) = \langle U + \be^{-1}\log\rho \rangle$ and the equilibrium free energy $
\caF^\text{eq} := \caF(\rho^\text{eq}) = -\be^{-1} \log Z$.\\

This formalism can directly be applied to statistical forces. Assuming
$\rho^\text{eq}_x(\eta) = \frac{1}{Z_x}\,e^{-\be U(x,\eta)}$, then we have
\begin{equation}
\begin{split}
f(x) &= -\langle \nabla_x U \rangle^x
= -\nabla_x \caF^\text{eq}(x) -
\frac{1}{\be}\,\Bigl\langle
\nabla_x \log\frac{\id\rho_x}{\id\rho^\text{eq}_x}
\Bigr\rangle^x
\\
&= -\nabla_x \caF^\text{neq}(x)
+ \frac{1}{2} \bigl\langle
\nabla_x \bigl[ \langle W \rangle_\eta^x - \langle W \rangle_{\rightarrow\eta}^x
\bigr] \bigr\rangle^x + O(\ve^3)
\end{split}
\end{equation}
or
\begin{equation}
f(x) \cdot \id x = -\id \caF^\text{neq} + \frac{1}{2}
\,[\, \dbar W(x) - \dbar W^-(x) ] + O(\ve^3)
\end{equation}
where $\dbar W(x) = \int \langle W \rangle^{x+\id x}_\eta\,\rho^x(\eta)\id\eta$ is the usual work of driving forces along the thermodynamic transformation
$x \mapsto x + \id x$, whereas
$\dbar W^-(x) = \int \langle W \rangle^{x+\id x}_{\rightarrow\eta}\,\rho^x(\eta)\id\eta$ can be interpreted as the work along a reversed \emph{spontaneous} process
$x + \id x \mapsto x$ (more precisely: within an ensemble interpretation it reads the work done along a typical excitation path realizing the empirical occupations given by $\rho_x$ in the steady state described by $\rho_{x+\id x}$).

\section{Conclusion}
The notions of thermodynamic landscape and potential are at the core of many thermodynamic considerations and heuristics.   For a probe in a nonequilibrium medium the work done by or on the probe may be path-dependent even under quasistatic conditions.  It allows for a greater phenomenology in statistical forces, avoiding gradient flow and enabling oscillatory motion.  That forces obtain a rotational component when induced by nonequilibrium media is not surprising, but it is  important and interesting to see the explicit nature and mechanism.

From response theory we have obtained a general characterization of how that picture arises. The rotational component can be expressed via the dependencies on the probe of the entropy flux and the frenesy in the nonequilibrium medium.  A certain ``twist'' between the entropic and the frenetic contributions induces the nongradient nature of the force.\\

\noindent{\bf Acknowledgment.\;} We thank Urna Basu for many discussions. KN acknowledges the support from the Grant Agency of the Czech Republic, grant no.~17-06716S.

\appendix

\section{Frenesy}\label{dac}
At various places in the main text and starting with equations \eqref{rnd}--\eqref{beg}, appears the frenesy $D$ of the medium, either as function of the coupling or as function of the probe position. We add here an explicit expression for $D$ for Markov jump processes; see also \cite{fren} for many more details.

Suppose for simplicity that the medium in contact with the static probe $x$ undergoes a Markov jump process $(\eta_s, 0\leq s\leq t)$ with transition rates
\begin{equation}\label{trr}
k(\eta,\eta') = \Phi_{x,\lambda}(\eta,\eta')\,\exp{[\beta U_\lambda(x,\eta)]}\,\, \exp {[\beta W(\eta,\eta')/2]}
\end{equation}
for symmetric prefactor $\Phi_{x,\lambda}(\eta,\eta') = \Phi_{x,\lambda}(\eta',\eta)$ and with work $W(\eta,\eta') = -W(\eta',\eta)$ done by the driving forces in the transition $\eta\rightarrow \eta'$.  
For $W\equiv 0$ there is (global) detailed balance with potential $U_\lambda$ at inverse temperature $\beta$.

Frenesy in \eqref{beg} is a (time-extensive) path-quantity $D=D_{x,\lambda,\text{driving}}(\omega)$, where $\omega$ is the (medium) $\eta$-trajectory during $[0,t]$, which complements the variable entropy flux
$S(\omega)$ and they both together uniquely determine the plausibility of a trajectory $\omega$. In particular,
$D(\omega) = D(\theta \omega)$ is time-symmetric with time-reversal $(\theta \omega)_s =\eta_{t-s}$.
The frenesy $D$ gets always expressed relative to some other process, changing some physical parameters in \eqref{trr}. Let us here however keep the nonequilibrium driving and the environment temperature fixed but we suppose changing  the probe position $x$ or the coupling $\lambda$.

There are two parts in $D$, one related to the log-reactivities,
\[
r(\eta,\eta') := -\log \Phi_{x,\lambda}(\eta,\eta') - \frac {\beta}{2}\,[ U_\lambda(x,\eta)+ U_\lambda(x,\eta')]
\]
and one obtained from the escape rate
\[
\xi(\eta) = \exp{[\beta U_\lambda(x,\eta)]}\,\sum_{\eta'} \Phi_{x,\lambda}(\eta,\eta')\, \exp [\beta W(\eta,\eta')/2].
\]
Then, the frenesy equals
\begin{equation}\label{daa}
D(\omega) = \sum_s r(\eta_{s^{-}},\eta_s) + \int_0^t\id s\,\xi(\eta_s)
\end{equation}
where the first sum is over the jump times in the trajectory $\omega$.

For Markov diffusion processes, a similar formula can be derived. 
The important and interesting point is that the variables $\eta$ respond to changes in the distribution through the frenesy $D$ and through the entropy flux $S$, which are explicit for any given model dynamics, as is very useful for the response formalism of Section \ref{resp}.

\section{Komatsu-Nakagawa formula}\label{knv}
Here we sketch here how the results in \cite{naokom} that we have used in Section \ref{notso} can be recovered within the framework of Section \ref{resp}.
The process $P_1$ now refers to equilibrium and the process $P_2$ to the nonequilibrium process.  We will however drop the subscripts $1,2$ and e.g. write $P^\text{eq}$ and $\rho^\text{eq}$ for the reference equilibrium path-space and stationary distributions.  Similarly, the $S_2-S_1=S_\text{i} = \be W$ is the associated entropy flux linear in the driving amplitude $\ve$, and $D_i=D_2-D_1$.  Then, as in \eqref{ppt} or via \cite{pp},
one finds for $\tau\uparrow \infty$,
\begin{equation}\label{bare}
\begin{split}
\frac{\id\rho}{\id\rho^\text{eq}}(\eta) &=
\Bigl\langle e^{S_\text{i}/2 - D_\text{i}}
\,\Bigl|\,\Bigr. \eta_\tau = \eta
\Bigr\rangle^\text{eq}
\\
&= 1 - \langle S_\text{i} \rangle^\text{eq}_\eta +
\langle D_\text{i} S_\text{i} \rangle^\text{eq}_\eta
+ O(\ve^3)
\end{split}
\end{equation}
Time-reversal allowed to change the conditioning in the future to specifying an initial condition.  The last line is however formal with entropy flux and excess frenesy over an infinite time-interval.  That formal expression can be renormalized by using excess quantities.
The excess entropy flux along relaxation to stationarity from an initial $\eta$ equals
\begin{equation}
\begin{split}
\langle S_\text{i} \rangle^\text{(ex)}_\eta &=
\langle S_\text{i} \rangle_\eta - \langle S_\text{i} \rangle
\\
&= \Bigl\langle \Bigl( 1 - D_\text{i} + \frac{S_\text{i}}{2} \Bigr)
S_\text{i} \Bigr\rangle^\text{eq}_\eta -
\Bigl\langle \frac{\rho}{\rho^\text{eq}}(\eta_0)
\Bigl( 1 - D_\text{i} + \frac{S_\text{i}}{2} \Bigr) S_\text{i}
\Bigr\rangle^\text{eq} + O(\ve^3)
\end{split}
\end{equation}
Substituting the first-order approximation \eqref{bare} for $\rho$, we get
\begin{equation}\label{ex-ent}
\langle S_\text{i} \rangle^\text{(ex)}_\eta =
\langle S_\text{i} \rangle^\text{eq}_\eta -
\langle D_\text{i} S_\text{i} \rangle^\text{eq}_\eta +
\Bigl\langle \frac{S_\text{i}^2}{2} \Bigr\rangle^\text{eq}_\eta -
\Bigl\langle \frac{S_\text{i}^2}{2} \Bigr\rangle^\text{eq} +
\Bigl\langle \Bigl(
\langle S_\text{i} \rangle^\text{eq}_{\eta'} \Bigr)^2
\Bigr\rangle^\text{eq} + O(\ve^3)
\end{equation}
By taking the expectation with respect to $\rho^\text{eq}(\eta)$ of \eqref{ex-ent} we check that the last (constant) term
\[
\Bigl\langle \Bigl(
\langle S_\text{i} \rangle^\text{eq}_{\eta'} \Bigr)^2
\Bigr\rangle^\text{eq} =  \left\langle \langle S_\text{i} \rangle^\text{(ex)}_\eta\right\rangle^\text{eq}
\]
coincides with the mean excess entropy flux when started from equilibrium.

Analogously as in~\eqref{ex-ent} we can calculate the excess entropy flux but now conditioned on ending at $\eta$.  A tedious but straightforward calculation gives
\begin{equation}\label{2ex}
\begin{split}
\langle S_\text{i} \rangle^\text{(ex)}_{\rightarrow \eta} &:=
\langle S_\text{i} \rel \eta_\tau = \eta \rangle -
\langle S_\text{i} \rangle\\
&= -\langle S_\text{i} \rangle^\text{eq}_\eta +
\langle D_\text{i} S_\text{i} \rangle^\text{eq}_\eta +
\Bigl\langle \frac{S_\text{i}^2}{2} \Bigr\rangle^\text{eq}_\eta -
\Bigl\langle \frac{S_\text{i}^2}{2} \Bigr\rangle^\text{eq} -
\Bigl( \langle S_\text{i} \rangle^\text{eq}_\eta \Bigr)^2
+ O(\ve^3)
\end{split}
\end{equation}
Hence, subtracting \eqref{2ex} from \eqref{ex-ent}, we get
\begin{equation}\label{sil}
\begin{split}
\langle S_\text{i} \rangle_\eta -
\langle S_\text{i} \rangle_{\rightarrow \eta} &=
2\langle S_\text{i} \rangle^\text{eq}_\eta -
2\langle D_\text{i} S_\text{i} \rangle^\text{eq}_\eta +
\Bigl( \langle S_\text{i} \rangle^\text{eq}_\eta \Bigr)^2 +
\bigl\langle \langle S_\text{i} \rangle^\text{(ex)}_{\eta'}
\bigr\rangle^\text{eq} +
O(\ve^3)
\end{split}
\end{equation}
We now compare with
\eqref{bare} and obtain
\begin{equation}\label{qual}
\frac{\id\rho}{\id\rho^\text{eq}}(\eta)
= 1 -  \frac 1{2}[\langle S_\text{i} \rangle_\eta -
\langle S_\text{i} \rangle_{\rightarrow \eta}] +
\frac 1{2}\Bigl( \langle S_\text{i} \rangle^\text{eq}_\eta \Bigr)^2 +
\frac 1{2}\bigl\langle \langle S_\text{i} \rangle^\text{(ex)}_{\eta'}
\bigr\rangle^\text{eq} +
O(\ve^3)
\end{equation}
Finally we take the logarithm of that expression and expand it to second order around equilibrium.  Observe that the only linear term in \eqref{qual} comes from the first term in the right-hand side of \eqref{sil}.
That delivers
immediately the Komatsu-Nakagawa formula \eqref{naoko} of \cite{naokom}, from which we start in Section \ref{notso}.

\end{document}